\documentclass [12pt] {article}
\usepackage {latexsym, amssymb}
\usepackage {amsmath}

\begin {document}

\title {Two Squares of Opposition: for Analytic and Synthetic Propositions}

\author {Andrew Schumann}
\maketitle
\begin {abstract}
In the paper I prove that there are two squares of opposition. The
unconventional one is built up for synthetic propositions. There
$a, i$ are contrary, $a, o$ (resp.\ $e, i$) are contradictory, $e,
o$ are subcontrary, $a,e$ (resp.\ $i, o$) are said to stand in the
subalternation.
\end {abstract}

\section{Introduction}

Since Aristotle the best known logical pattern has been presented
by square of opposition. However, since Kant and Rickert many
philosophers have paid attention that this square does not satisfy
synthetic propositions. In this paper we are proving that there
are two squares of opposition under following assumptions:
\begin{itemize}
    \item we have the Boolean complement;
    \item synthetic propositions cannot be reduced to manipulations with Venn diagrams, because they do not suppose including relations.
\end{itemize}

\section{The short history of the square of opposition}

Aristotle proposed the following four oppositions: contradiction,
contrariety, relation, and privation (\emph{Categories}, Chapter
10, \emph{Metaphysics}, Book I) that semantically underlay the
square of opposition:
\begin{quote}We must next explain the various senses in which the term
`opposite' is used. Things are said to be opposed in four senses:
(i) as correlatives to one another, (ii) as contraries to one
another, (iii) as privatives to positives, (iv) as affirmatives to
negatives (\emph{Cathegories}, 10).\end{quote}

Aristotle himself described relations of the square of opposition
to represent singular expressions (\emph{Prior Analytics}, Chapter
46), see figure 1. He had an intuition that quantifiers (both
universal and existential ones) satisfy the semantic relations of
the square, too:
\begin{quote}An affirmation is opposed to a denial in the sense which I
denote by the term `contradictory', when, while the subject
remains the same, the affirmation is of universal character and
the denial is not. The affirmation `every man is white' is the
contradictory of the denial `not every man is white', or again,
the proposition `no man is white' is the contradictory of the
proposition `some men are white'. But propositions are opposed as
contraries when both the affirmation and the denial are universal,
as in the sentences `every man is white', `no man is white',
`every man is just', `no man is just (\emph{On Interpretation},
7).\end{quote}

\begin{center}
\begin{figure}[ht] \unitlength1.5cm
\begin{picture}(3.0,3.0)
\put(0,0){\line(0,2){2.0}}
\put(0,2){\line(2,0){2.0}}\put(2,0){\line(0,2){2.0}}\put(0,0){\line(2,0){2.0}}
\put(0,2.2){$S \text{ is } P$}\put(2.1,2.2){$S \text{ isn't not-}
P$} \put(0,0.1){$S \text{ is not-}P$}\put(2.1,0.1){$S \text{ is
not }P$}
\end{picture}
\caption{Aristotle's square of opposition.}
\label{fig:schu1}\end{figure}
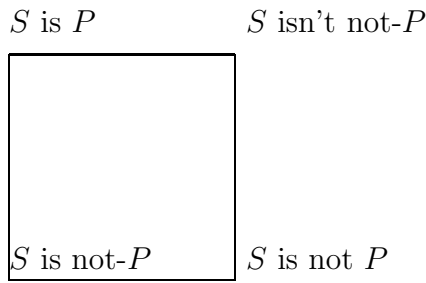
\end{center}

However, for the first time, Apuleius explicitly claimed that
quantified propositions satisfy the square. He wrote a short book,
the \emph{Peri Hermeneias} \cite{Apul2}, about logic that was used
for centuries in teaching. This book is the most famous in the
history of logic for including the first appearance of the square
of opposition, the best known logical schema for the pedagogic
purpose \cite{Apul2}. He considered the four oppositions:
contrary, subcontrary, contradictory, subalternation. First, he
described that the two \emph{incongruae} (contrary) propositions,
on the left and right sides of the top of the square, never can be
true at the same time and nonetheless are sometimes false at the
same time. For example, when some pleasures are good, both
universal propositions are false at the same time, since it is
impossible that every pleasure is both a good and not a good. The
two propositions along the bottom line (i.e.\ the mirror-image of
the contrary) are called \emph{subpares} (subcontrary). They are
never false at the same time, but they can be true at the same
time. Hence, to confirm that some pleasure is a good we cannot use
an argument that some other pleasure is not a good. Further, we
pair together the \emph{alterutrae} (contradictory) propositions,
if we add a negation to each of the pair of alternates, e.g.\ not
every pleasure is a good means that some pleasure is not a good.
Finally, the \emph{subalternation} appears between universal and
particular propositions, when the universal implies the
particular, e.g.\ if every pleasure is a good, then some pleasure
is a good.

The meaning of propositions that satisfy the square of opposition
can be checked on Venn diagrams. Recall that a Venn diagram is an
ellipse that designates an extent of a concept $A$, i.e.\ a class
of all real things that are denoted by $A$. These things are
called denotations. By assumption, all inner points of ellipse
designate appropriate real things. For instance, `every man is
mortal' is a true proposition, because the Venn diagram of `man'
is included into the Venn diagram of `the mortal being' (i.e.\ all
denotations of `man' occur among denotations of `the mortal
being').

Kant first paid attention that there exists a true universal
proposition like `all bodies are heavy' such that Venn diagrams of
its subject and predicate do not assume the including relation.
So, the Venn diagram of `body' just intersects the Venn diagram of
`heavy':

\begin{quote}In all judgments in which the relation of a subject to the
predicate is thought \dots, this relation is possible in two
different ways. Either the predicate to the subject $A$, as
something which is (covertly) contained in this concept $A$; or
outside the concept $A$, although it does indeed stand in
connection with it. In the one case I entitle the judgment
analytic, in the other synthetic. Analytic judgments (affirmative)
are therefore those in which the connection of the predicate with
the subject is thought through identity; those in which this
connection is thought without identity should be entitled
synthetic (\dots) If I say, for instance, `\emph{All bodies are
extended}', this is an analytic judgment. For I do not require to
go beyond the concept which I connect with `\emph{body}' in order
to find extension as bound up with it \dots The judgment is
therefore analytic. But when I say, `\emph{All bodies are heavy}',
the predicate is something quite different from anything that I
think in the mere concept of body in general; and the addition of
such a predicate therefore yields a synthetic judgment.

Judgments of experience, as such, are one and all synthetic. For
it would be absurd to found an analytic judgment on experience.
Since, in framing the judgment, I must not go outside my concept,
there is no need to appeal to the testimony of experience in its
support \cite{Kant}.\end{quote}

Thus, in Kant's opinion, only analytic judgments satisfy the
square of opposition. For synthetic judgments Venn diagrams lose
any sense and, as a result, we cannot apply the square for them.
In Aristotelian logic there is the inverse relation between
content (class of all connotations) and extent (class of all
denotations) of a concept. By continuing the Kant's ideas,
Heinrich Rickert claimed that \emph{for synthetic judgments
(propositions) there is the direct relation between content and
extent of a concept}. Therefore we cannot use Venn diagrams there
at all.

Thus, according to Kant and Rickert, there are two logics (the
Aristotelian for analytic propositions, where we can use Venn
diagrams and the square of opposition, and the non-Aristotelian
for synthetic propositions without Venn diagrams manipulations).
This distinction entails another distinction (proposed first by
Wilhelm Windelband and Heinrich Rickert) between two kinds of
sciences: natural sciences (\emph{Naturwissenschaften}) and
cultural sciences (\emph{Geisteswissenschaften}). In the first the
Aristotelian logic is used by applying a nomothetic approach, in
the second the non-Aristotelian by applying an idiographic
approach. The idiographic approach is concerned with individual
phenomena, as in biography and much of history, while its
opposite, the nomothetic approach, aims to formulate laws as
general propositions.

In the history as in an individualizing science (\emph{eine
individualisierende Wissenschaft}) we obtain the direct relation
between content and extent of historical concepts: the more
general historical concept is more value relevant at the same
time:
\begin{quote}It obviously remains true that in characterizing one
part of historical activity, we can speak of a re-creative
understanding of the ``spiritual'' [der geistigen] world. But the
concepts of understanding and re-creation are also too imprecise
and general to provide a fully autonomous and exhaustive
characterization of the nature of all historical representation.
As regards understanding, it is important, first, that the object
of understanding in history is always something more than merely
real; namely, it is value relevant and meaningful. And second, to
remain within the domain of history, the value relevant and the
meaningful are comprehended not in a generalizing fashion but in
an individualizing fashion, even though their content may be only
relatively historical. Finally, even the concept of the
``re-creation'' of historical individuality acquires its precise
significance for the theory of the historical sciences only on the
basis of the concept of the individualizing understanding of
meaning \cite{Rickert}.\end{quote}

\begin{center}
\begin{figure}[ht] \unitlength1.7cm
\begin{picture}(4.0,4.0)
\put(0,0){\line(0,4){4.0}}
\put(0,4){\line(4,0){4.0}}\put(4,0){\line(0,4){4.0}}\put(0,0){\line(4,0){4.0}}
\put(0,4.2){\emph{Any goodness will be
crowned}}\put(4.1,4.2){\emph{Some goodness will be crowned}}
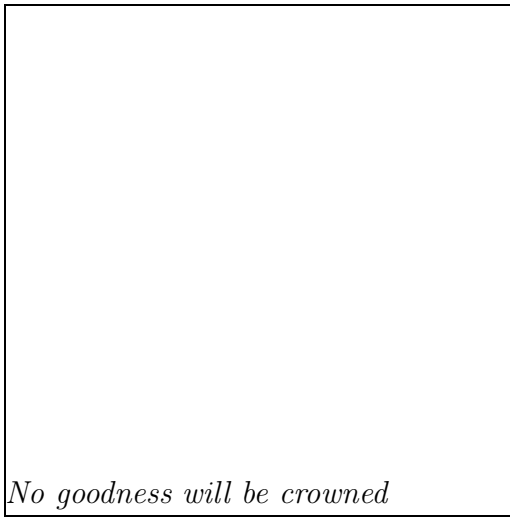
\put(0,0.1){\emph{No goodness will be
crowned}}\put(4.1,0.1){\emph{Some goodness will not be crowned}}
\end{picture}
\caption{An example of the square of opposition for
individualizing propositions.} \label{fig:schu1}\end{figure}
\end{center}

Under these conditions, the general does not imply the particular.
For instance, as we saw, we cannot differ the universal
proposition `all bodies are heavy' from the particular one `some
bodies are heavy' by Venn diagrams, because the extents of their
concepts (the extents of `bodies' and `heavy') are just
intersected.

Rickert did not think of creating a new square of opposition that
may become suitable for describing semantic oppositions between
synthetic (historical, individualizing) propositions. If we set up
such a problem, we will start in distinguishing between general
and particular synthetic propositions.

For analytic propositions while we move from the general (i.e.\
the concept with the larger extent and the smaller content) to the
particular (i.e.\ the concept with the smaller extent and the
larger content), we are losing definiteness and certainty. For
synthetic propositions, the general and the particular are two
different points of view, because both have different extents and
different contents with the same certainty, which satisfy a direct
relation between them.

In the Apuleian square of opposition there is a duality between
the general and the particular. Indeed, for the general there is a
contrary negation and for the particular a subcontrary negation,
thereby the contrary negation tends to be maximized and the
subcontrary negation tends to be minimized. In the new square of
opposition (see figure 2) we could propose another \emph{duality
that takes place between the affirmation and the negation}. In
this case the contrary negation holds between the general
affirmative proposition and the particular affirmative proposition
and the subcontrary negation between the general negative
proposition and the particular negative proposition.

In next sections we are formally proving that there exist two
squares of opposition and, correspondingly, two syllogistics (the
first for analytic propositions and the second for synthetic
propositions).

\section{Synthetic syllogictics}

In Aristotle's syllogistics, analytic propositions in Kant's words
are formalized. In such propositions, a subject is thought within
a predicate, the more common concept, for example: `Socrates is a
\emph{man}' or `All people are \emph{animals}'. Therefore the
predicate in formulas $S \mathbf{a} P$ (`every $S$ is $P$'), $S
\mathbf{i} P$ (`some $S$ are $P$'), $S \mathbf{e} P$ (`no $S$ is
$P$'), $S \mathbf{o} P$ (`some $S$ are not $P$') may be replaced
by nouns, but not by adjectives. In other words, we have the
following grammar: `noun 1 (subject) + is + noun 2 (predicate)',
and the concept of `noun 1' is a kind (particular) of the concept
of `noun 2', i.e. `noun 2' is thought as general for `noun 1'.

However, how far can we consider propositions like `Socrates is
white', `All bodies are heavy' within a conventional syllogistics
formalizing just analytic propositions? At the first blush, these
troubles might be deleted if we transformed an appropriate
adjective into a noun. For example, the proposition `Socrates is
white' may be converted to the proposition `Socrates is a white
being', and `All bodies are heavy' to `All bodies are something
heavy'. However, such a transformation does not solve our problem,
because the predicate is not general for the subject still. Thus,
Socrates' whiteness is not his substantial attribute and the
general of bodies is space, but not weight. Any body is thought
first as space entity.

For the first time, Aristotle noticed that there are propositions
that have been called synthetic since Kant and these propositions
cannot be used in syllogistics. Aristotle's counterexample was as
follows: ``He who sits, writes and Socrates is sitting, then
Socrates is writing'' (\emph{Topics} 10). This wrong syllogism was
caused by using synthetic propositions. Kant's key example of
synthetic propositions: `All bodies are heavy'. They have the
following grammar: `noun 1 (subject) + is + adjective
(attribute)'. The connective `\dots is \dots' of synthetic
propositions is understood in this paper as follows: $A \,\,is\,\,
B \equiv (\exists C (C \,\,is\,\, A) \wedge \forall C \forall D
((C \,\,is\,\, A \wedge D \,\,is\,\, A) \Rightarrow C \,\,is\,\,
D) \wedge \forall C (C \,\,is\,\, A \wedge C \,\,is\,\, B))$. This
understanding corresponds to the Kantian-Rickertian approach.

Assume that all the syllogistic synthetic propositions have the
following meaning:
\begin{itemize}
    \item `All $S$ are $P$' (`All bodies are heavy'): there
    exist $A$ such that $A$ is $S$ or for any $A$, $A$ is
    $S$ and $A$ is $P$;
    \item `Some $S$ are $P$' (`Socrates is white'): there exist
    $A$ such that `$A$ is $S$' is false, but `$A$ is
    $P$' is true;
    \item `No $S$ are $P$' (`No bodies are angels'): there
    exist $A$ such that `$A$ is $P$' is false or `$A$ is
    $S$' is true;
    \item `Some $S$ are not $P$' (`Socrates is not black'): for
    any $A$, `$A$ is $S$' is false and there exist $A$ such
    that `$A$ is $P$' is false or `$A$ is
    $S$' is false.
\end{itemize}

Let us propose now the syllogistic system formalizing synthetic
propositions. This system is said to be \emph{synthetic
syllogistics}, while we are assuming that Aristotelian syllogistic
is analytic. The basic logical connectives of synthetic
syllogistic are as follows: $\mathfrak{a}$ (`every + noun + is +
adjective'), $\mathfrak{i}$ (`some + noun + is + adjective'),
$\mathfrak{e}$ (`no + noun + is + adjective') and $\mathfrak{o}$
(`some + noun + is not + adjective') that are defined in synthetic
ontology in the following way:
\begin {equation}S\mathfrak{a} P :=(\exists A (A\,\,is\,\, S) \vee (\forall A (A\,\,is\,\, P \wedge A
\,\,is\,\, S)));\end {equation}
\begin {equation}S\mathfrak{i} P := \forall A (A\,\,is\,\, P \wedge \neg ( A
\,\,is\,\, S));\end {equation}
\begin {eqnarray}S\mathfrak{o} P := \neg(\exists A (A\,\,is\,\, S) \vee (\forall A (A\,\,is\,\, P \wedge A
\,\,is\,\, S))), \,\,i.e. {}\nonumber\\(\forall A \neg(A\,\,is\,\,
S) \wedge \exists A (\neg (A\,\,is\,\, P) \vee \neg (A \,\,is\,\,
S)));\end {eqnarray}
\begin {equation}S\mathfrak{e} P := \neg\forall A (A\,\,is\,\, P \wedge \neg ( A
\,\,is\,\, S)), \,\,i.e.\,\, \exists A (\neg (A\,\,is\,\, P) \vee
(A \,\,is\,\, S)).\end {equation}

Now let us formulate axioms of synthetic syllogistics:

\begin {equation}S \mathfrak{a} P \Rightarrow S \mathfrak{e} P;\end {equation}
\begin {equation}S \mathfrak{o} P \Rightarrow P \mathfrak{o} S;\end {equation}
\begin {equation}(M \mathfrak{a} P \wedge S \mathfrak{a} M) \Rightarrow S \mathfrak{a} P;\end {equation}
\begin {equation}(M \mathfrak{a} P \wedge S \mathfrak{e} M) \Rightarrow S \mathfrak{e}
  P.\end {equation}

In synthetic syllogistics we have a novel square of opposition
that we call the \emph{synthetic square of opposition} (see figure
3), where the following theorems are inferred: $S \mathfrak{a} P
\Rightarrow \neg (S \mathfrak{o} P)$, $\neg (S \mathfrak{o} P)
\Rightarrow S \mathfrak{a} P$, $S \mathfrak{i} P \Rightarrow \neg
(S \mathfrak{e} P)$, $\neg (S \mathfrak{e} P) \Rightarrow S
\mathfrak{i} P$, $S \mathfrak{e} P \Rightarrow \neg (S
\mathfrak{i} P)$, $\neg (S \mathfrak{i} P) \Rightarrow S
\mathfrak{e} P$, $S \mathfrak{o} P \Rightarrow  \neg (S
\mathfrak{a} P)$, $\neg (S \mathfrak{a} P) \Rightarrow S
\mathfrak{o} P$, $S \mathfrak{a} P \Rightarrow \neg (S
\mathfrak{i} P)$, $S \mathfrak{i} P \Rightarrow \neg (S
\mathfrak{a} P)$, $\neg (S \mathfrak{e} P) \Rightarrow S
\mathfrak{o} P$, $\neg (S \mathfrak{o} P) \Rightarrow S
\mathfrak{e} P$, $S \mathfrak{a} P \Rightarrow S \mathfrak{e} P$,
$S \mathfrak{i} P \Rightarrow S \mathfrak{o} P$, $S \mathfrak{e} P
\vee  S \mathfrak{i} P$, $\neg (S \mathfrak{e} P \wedge  S
\mathfrak{i} P)$, $S \mathfrak{a} P \vee S \mathfrak{o} P$, $\neg
(S \mathfrak{a} P \wedge S \mathfrak{o} P)$, $\neg (S \mathfrak{a}
P \wedge S \mathfrak{i} P)$, $S \mathfrak{e} P \vee S \mathfrak{o}
P$.

\begin{center}
\begin{figure}[ht] \unitlength1.5cm
\begin{picture}(3.0,3.0)
\put(0,0){\line(0,2){2.0}}
\put(0,2){\line(2,0){2.0}}\put(2,0){\line(0,2){2.0}}\put(0,0){\line(2,0){2.0}}
\put(0,2.2){$S \mathfrak{a} P$}\put(2.1,2.2){$S \mathfrak{i} P$}
\put(0,0.1){$S \mathfrak{e} P$}\put(2.1,0.1){$S \mathfrak{o} P$}
\end{picture}
\caption{The synthetic square of opposition (for synthetic
syllogistics, where synthetic  propositions are formalized).}
\label{fig:schu3}\end{figure}
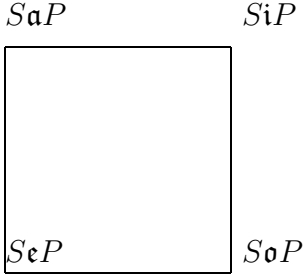
\end{center}

\section{Non-Archimedean models of Aristotelian syllogistics and
synthetic syllogistics}

Suppose $B$ is a complete Boolean algebra with the bottom element
0 and the top element 1 such that the cardinality of its domain $|
B|$ is an infinite number. Build up the set $B^B$ of all functions
$f \colon B \mapsto B$. The set of all complements for finite
subsets of $B$ is a filter and it is called a Frech\'{e}t filter,
it is denoted by $\mathcal{U}$. Further, define a new relation
$\approx$ on the set $B^B$ by $f \approx g = \{a \in B \colon f(a)
= g(a)\} \in \mathcal{U}$. It is easily proved that the relation
$\approx$ is an equivalence. For each $f \in B^B$ let $[f]$ denote
the equivalence class of $f$ under $\approx$. The ultrapower
$B^B/\mathcal{U}$ is then defined to be the set of all equivalence
classes $[f]$ as $f$ ranges over $B^B$. This ultrapower is called
a \textit{nonstandard} (or \textit{non-Archimedean})
\textit{extension of Boolean algebra} $B$, for more details see
\cite{Robinson} and \cite{Schumann1}. It is denoted by ${}^*B$.

There exist two groups of members of ${}^*B$: (1) functions that
are constant, e.g.\ $f(a) = m \in B$ on the set $\mathcal{U}$, a
constant function $[f = m]$ is denoted by ${}^*m$, (2) functions
that are not constant. The set of all constant functions of
${}^*B$ is called standard set and it is denoted by ${}^\circ B$.
The members of ${}^\circ B$ are called standard. It is readily
seen that $B$ and ${}^\circ B$ are isomorphic.

We can extend the usual partial order structure on $B$ to a
partial order structure on ${}^\circ B$:

\begin{enumerate}
\item  for any members $x, y \in B$ we have $x \leq y$ in $B$ iff
${}^*x \leq {}^*y$ in ${}^\circ B$,

\item  each member ${}^*x \in {}^\circ B \backslash \{{}^*0\}$
(i.e. that is not a bottom element ${}^*0$ of ${}^\circ B$) is
greater than any number $[f] \in {}^*B\backslash {}^\circ B$,
i.e.\ ${}^*x
> [f]$ for any $x \in B$, where $[f]$ is not constant function, \item ${}^*0$ is the bottom element of ${}^*B$.
\end{enumerate}

Notice that under these conditions, there exist the top element
${}^ *1 \in {}^*B$ such that $1 \in B$ and the bottom element
${}^*0 \in {}^*B$ such that $0 \in B$.

The ordering conditions mentioned above have the following
informal sense: (1) the sets ${}^\circ B$ and $B$ have isomorphic
order structure; (2) the set ${}^*B\backslash \{{}^*0\}$ contains
actual infinities that are less than any member of ${}^\circ
B\backslash \{{}^*0\}$. These members are called \emph{Boolean
infinitesimals}.

Introduce three operations `$\sup$', `$\inf$', `$\neg$' in the
partial order structure of ${}^*B$:
$$\inf([f], [g]) = [\inf(f, g)];$$
$$\sup([f], [g]) = [\sup(f, g)];$$
$$\neg[f] = [\neg f].$$
This means that a nonstandard extension ${}^*B$ of a Boolean
algebra $B$ preserves the least upper bound `$\sup$', the greatest
lower bound `$\inf$', and the complement `$\neg$' of $B$.

Consider the member $[h]$ of ${}^*B$ such that $\{a \in B \colon
h(a) = f(\neg a)\} \in \mathcal{U}$. Denote $[h]$ by $[f\neg]$.
Then we see that $\inf([f], [f\neg]) \geq {}^*0$ and $\sup([f],
[f\neg]) \leq {}^*1$. Really, we have several cases.

\begin{enumerate}
\item  \textit{Case 1}. The members $\neg[f]$ and $[f\neg]$ are
incompatible. Then $\inf([f]$, $[f\neg]) \geq {}^*0$ and
$\sup([f]$, $[f\neg]) \leq {}^*1$,

\item  \textit{Case 2}. Suppose $\neg[f] \geq [f\neg]$. In this
case $\inf([f], [f\neg]) = {}^*0$ and $\sup([f], [f\neg]) \leq
{}^*1$.

\item  \textit{Case 3}. Suppose $\neg [f] \leq [f\neg]$. In this
case $\inf([f], [f\neg]) \geq {}^*0$ and $\sup([f], [f\neg]) =
{}^*1$.

\item  \textit{Case 4}. The members $[f]$ and $\neg[f\neg]$ are
incompatible. Then $\inf(\neg[f]$, $\neg [f\neg]) \geq {}^*0$ and
$\sup(\neg [f], \neg[f\neg]) \leq {}^*1$,

\item  \textit{Case 5}. Suppose $\neg[f\neg] \geq [f]$. In this
case $\inf(\neg [f], \neg [f\neg]) \geq {}^*0$ and $\sup(\neg [f],
\neg [f\neg]) = {}^*1$.

\item  \textit{Case 6}. Suppose $\neg [f\neg] \leq [f]$. In this
case $\inf(\neg [f], \neg [f\neg]) = {}^*0$ and $\sup(\neg [f],
\neg [f\neg]) \leq {}^*1$.

\item  \textit{Case 7}. The members $\neg[f\neg]$ and $\neg[f]$
are incompatible. Then $\inf([f]$, $\neg[f\neg]) \geq {}^*0$ and
$\sup([f], \neg[f\neg]) \leq {}^*1$,

\item  \textit{Case 8}. Suppose $\neg[f] \geq \neg[f\neg]$. In
this case $\inf([f], \neg[f\neg]) = {}^*0$ and $\sup([f],
\neg[f\neg]) \leq {}^*1$.

\item  \textit{Case 9}. Suppose $\neg [f] \leq \neg[f\neg]$. In
this case $\inf([f], \neg[f\neg]) \geq {}^*0$ and $\sup([f],
\neg[f\neg]) = {}^*1$.

\item  \textit{Case 10}. The members $[f]$ and $[f\neg]$ are
incompatible. Then $\inf(\neg[f]$, $[f\neg]) \geq {}^*0$ and
$\sup(\neg [f], [f\neg]) \leq {}^*1$,

\item  \textit{Case 11}. Suppose $[f\neg] \geq [f]$. In this case
$\inf(\neg [f], [f\neg]) \geq {}^*0$ and $\sup(\neg [f], [f\neg])
= {}^*1$.

\item  \textit{Case 12}. Suppose $[f\neg] \leq [f]$. In this case
$\inf(\neg [f], [f\neg]) = {}^*0$ and $\sup(\neg [f], [f\neg])
\leq {}^*1$.
\end{enumerate}

\newtheorem{definitions}{Definition}\begin{definitions} Now define hyperrational valued matrix
logic $\mathfrak M_B$ as the ordered system $\langle {}^*B,
\{{}^*1\}, \neg, \Rightarrow, \vee, \wedge \rangle$, where

\begin{enumerate}
\item  ${}^*B$ is the set of truth values,

\item  $\{{}^*1\}$ is the set of designated truth values,

\item  for all $[x] \in {}^*B$, $\neg [x] = {}^*1 - [x]$,

\item  for all $[x], [y] \in {}^*B$, $[x] \Rightarrow [y] = {}^*1
- \sup([x], [y]) + [y]$,

\item  for all $[x], [y] \in {}^*B$, $[x] \wedge [y] = \inf([x],
[y])$,

\item  for all $[x], [y] \in {}^*B$, $[x] \vee [y] = \sup([x],
[y])$.
\end{enumerate}
\end{definitions}

\newtheorem{propositions}{Proposition}\begin{propositions}In $\mathfrak M_B$ there are only two squares of opposition.\end{propositions}
\emph{Proof}. We have just eight cases: (1) $[f] \leq [f \neg]$,
(2) $[f] \leq \neg [f \neg]$, (3) $[f \neg] \leq [f]$, (4) $[f
\neg] \leq \neg [f]$, (5) $\neg [f \neg] \leq [f]$, (6) $\neg [f
\neg] \leq \neg [f]$, (7) $\neg [f] \leq [f \neg]$, (8) $\neg [f]
\leq \neg [f \neg]$. Taking into account that couples $[f]$ and
$\neg [f]$ ($[f \neg]$ and $\neg [f \neg]$) are contradictory, we
can claim that there exist two squares of opposition:

\begin{itemize}
    \item if $[f \neg] \leq [f]$ (resp.\ $\neg [f] \leq \neg [f \neg]$), we have the conventional square of
opposition (see figure 4); if $\neg[f \neg] \leq [f ]$ (resp.\
$\neg [f] \leq [f \neg]$), we have its dual without changing
meaning;
    \item if $[f] \leq \neg [f \neg]$ (resp.\ $[f
\neg] \leq \neg [f]$), we have the synthetic square of opposition
(see figure 5); if $\neg [f\neg] \leq [f]$ (resp.\ $\neg[f] \leq
[f \neg]$), we have its dual without changing meaning.\hfill$\Box$
\end{itemize}

\begin{center}
\begin{figure}[ht] \unitlength1.5cm
\begin{picture}(3.0,3.0)

\put(0,0){\line(0,2){2.0}}
\put(0,2){\line(2,0){2.0}}\put(2,0){\line(0,2){2.0}}\put(0,0){\line(2,0){2.0}}
\put(0,2.2){$[f]$}\put(2.1,2.2){$[f\neg]$}
\put(0,0.1){$\neg[f\neg]$}\put(2.1,0.1){$\neg[f]$}

\end{picture}
\caption{In case $[f\neg] \leq \neg[f]$, the square of oppositions
for any members $[f]$, $[f\neg]$, $\neg[f]$, $\neg[f\neg]$ of
${}^*B$ holds true, i.e.\ $[f], [f\neg]$ are contrary, $[f],
\neg[f]$ (resp.\ $\neg[f\neg], [f\neg]$) are contradictory,
$\neg[f\neg], \neg[f]$ are subcontrary, $[f],\neg [f\neg]$ (resp.\
$[f\neg], \neg[f]$) are said to stand in the subalternation.}
\label{fig:schu5}\end{figure}
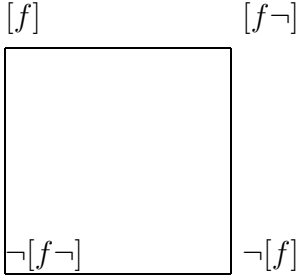
\end{center}
\begin{center}
\begin{figure}[ht] \unitlength1.5cm
\begin{picture}(3.0,3.0)
\put(0,0){\line(0,2){2.0}}
\put(0,2){\line(2,0){2.0}}\put(2,0){\line(0,2){2.0}}\put(0,0){\line(2,0){2.0}}
\put(0,2.2){$[f]$}\put(2.1,2.2){$\neg[f\neg]$}
\put(0,0.1){$[f\neg]$}\put(2.1,0.1){$\neg[f]$}
\end{picture}
\caption{In case $\neg[f\neg] \leq \neg[f]$, the \emph{synthetic}
square of oppositions for any members $[f]$, $[f\neg]$, $\neg[f]$,
$\neg[f\neg]$ of ${}^*B$ holds true, i.e.\ $[f], \neg[f\neg]$ are
contrary, $[f], \neg[f]$ (resp.\ $\neg[f\neg], [f\neg]$) are
contradictory, $\neg[f], [f\neg]$ are subcontrary, $[f], [f\neg]$
(resp.\ $\neg[f\neg], \neg[f]$) are said to stand in the
subalternation.} \label{fig:schu6}\end{figure}
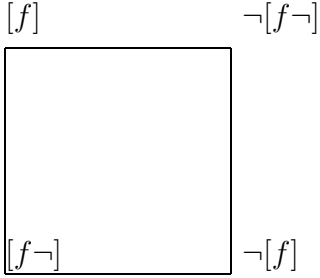
\end{center}

Now we can build models for atomic syllogistic formulas (i.e.
syllogistic formulas without propositional connectives) due to
algebra $\mathfrak M_B$.
\begin{definitions}
A structure $\mathfrak{B}=\langle O$, $I$, $\mathbf{\dot a}$,
$\mathbf{\dot e}$, $\mathbf {\dot i}$, $\mathbf {\dot o}$,
$\mathfrak {\dot a}$, $\mathfrak {\dot e}$, $\mathfrak {\dot i}$,
$\mathfrak {\dot o} \rangle$ is a \emph{non-Archimedean
syllogistic model} iff:
\begin{enumerate}
\setlength{\itemsep}{0pt} \setlength{\parsep}{0pt} \item $O$ is a
restriction of the set $\mathfrak M_B$ to an appropriate square
(triangle) of opposition (thereby the conventional square of
opposition should hold true for Aristotelian syllogistics and the
synthetic square of opposition holds for synthetic syllogistics).
\item $I$ is a mapping that associates a class of equivalence $
[f]\in O$ with each atomic syllogistic formula $S \diamond P$,
where $\diamond\in\{\mathbf a$, $\mathbf e$, $\mathbf i$, $\mathbf
o$, $\mathfrak a$, $\mathfrak e$, $\mathfrak i$, $\mathfrak o\}$,
so that $I(S \diamond P)= |S| {\dot \diamond} |P|$, where ${\dot
\diamond}\in\{\mathbf{\dot a}$, $\mathbf{\dot e}$, $\mathbf {\dot
i}$, $\mathbf {\dot o}$, $\mathfrak {\dot a}$, $\mathfrak {\dot
e}$, $\mathfrak {\dot i}$, $\mathfrak {\dot o}\}$ and
\begin{itemize}
    \item $|S| \mathbf{\dot a} |P| = [f]$ (resp.\ $|S| \mathbf{\dot a} |P| = \neg[f
    \neg]$);
    \item $|S| \mathbf{\dot e} |P| = [f\neg]$ (resp.\ $|S| \mathbf{\dot e} |P| =
    \neg[f]$);
    \item $|S| \mathbf{\dot i} |P| = \neg[f\neg]$ (resp.\ $|S| \mathbf{\dot i} |P| =
    [f]$);
    \item $|S| \mathbf{\dot o} |P| = \neg[f]$ (resp.\ $|S| \mathbf{\dot o} |P| = [f
    \neg]$);
    \item $|S| \mathfrak{\dot a} |P| = [f]$ (resp.\ $|S| \mathfrak{\dot a} |P| = \neg[f
    \neg]$);
    \item $|S| \mathfrak{\dot e} |P| = [f\neg]$ (resp.\ $|S| \mathfrak{\dot e} |P| =
    \neg[f]$);
    \item $|S| \mathfrak{\dot i} |P| = \neg[f\neg]$ (resp.\ $|S| \mathfrak{\dot i} |P| =
    [f]$);
    \item $|S| \mathfrak{\dot o} |P| = \neg[f]$ (resp.\ $|S| \mathfrak{\dot o} |P| = [f
    \neg]$).
    \end{itemize}    \end{enumerate}
\end{definitions}

We now give the truth conditions of Boolean combinations of atomic
syllogistic formulas in a non-Archimedean syllogistic model:
\begin{definitions}\mbox{}
\[
\begin{array}{lcl}
\mathfrak{B}\vDash\neg\phi & \text{iff} & \mathfrak{B}\nvDash\phi \\
\mathfrak{B}\vDash\phi\wedge\psi & \text{iff} & \mathfrak{B}\vDash\phi\text{ and }\mathfrak{B}\vDash\psi \\
\mathfrak{B}\vDash\phi\vee\psi & \text{iff} & \mathfrak{B}\vDash\phi\text{ or }\mathfrak{B}\vDash\psi \\
\mathfrak{B}\vDash\phi\Rightarrow\psi & \text{iff} &
\mathfrak{B}\vDash\neg\phi\text{ or }\mathfrak{B}\vDash\psi
\end{array}
\]
\end{definitions}

\section{Conclusion}

In this paper using non-Archimedean models I have just proved that
there are only two squares of opposition if we assume Boolean
algebra as the basis of an appropriate non-Archimedean extension.
The conventional square of opposition may be aimed for getting
analytic syllogistics (Aristotelian syllogistics) and the new one
for getting synthetic syllogistics (syllogistics, proposed in this
paper).

Andrew Schumann

Department of Philosophy and Science Methodology,

Belarusian State University, Minsk, Belarus

e-mail: Andrew.Schumann@gmail.com

\end {document}